\documentclass{acmsiggraph}               % final
\usepackage{mathptmx}
\usepackage{graphicx}
\usepackage{rotating}
\usepackage{enumerate}
\usepackage{microtype}

\usepackage{parskip}

\onlineid{0}

\acmformat{print}

\title{Two Dimensional Hidden Surface Removal\\ with Frame-to-frame
Coherence}

\author{John G. Whitington\thanks{e-mail: john@coherentgraphics.co.uk}\\
Coherent Graphics Ltd}


\begin{document}

%\teaser{
%  \centerline{\includegraphics[width=8cm]{filter-examples}}
%  \caption{Filters}
%}


\maketitle

%% Abstract section.

\begin{abstract} We describe a hidden surface removal algorithm for
  two-dimensional layered scenes built from arbitrary primitives, particularly
  suited to interaction and animation in rich scenes (for example, in
  illustration). The method makes use of a set-based raster representation to
  implement a front-to-back rendering model which analyses and dramatically
  reduces the amount of rasterization and composition required to render a
  scene. The method is extended to add frame-to-frame coherence analysis and
  caching for interactive or animated scenes. A powerful system of
  primitive-combiners called \textit{filters} is described, which preserves the
  efficiencies of the algorithm in highly complicated scenes.  The set
  representation is extended to solve the problem of correlated mattes, leading
  to an efficient solution for high quality antialiasing. A prototype implementation has been prepared.

%Citations can be done this way~\cite{Jobs95} or this more concise 
%way~\shortcite{Jobs95}, depending upon the application.

\end{abstract}

%% ACM Computing Review (CR) categories. 
%% See <http://www.acm.org/class/1998/> for details.
%% The ``\CRcat'' command takes four arguments.

\begin{CRcatlist}
%  \CRcat{K.6.1}{Management of Computing and Information Systems}%
%{Project and People Management}{Life Cycle};
%  \CRcat{K.7.m}{The Computing Profession}{Miscellaneous}{Ethics}
\CRcat{I.3.3}{Computer Graphics}{Picture/Image Generation}{Display algorithms}
\CRcat{I.3.3}{Computer Graphics}{Picture/Image Generation}{Antialiasing}
\CRcat{I.3.6}{Computer Graphics}{Methodology and Techniques}{Graphics data
structures and data types}
\end{CRcatlist}

%% The ``\keywordlist'' command prints out the keywords.
\keywordlist

%% The ``\copyrightspace'' command must be the first command after the 
%% start of the first section of the body of your paper. It ensures the
%% copyright space is left at the bottom of the first column on the first
%% page of your paper.

\vspace{2mm}
\copyrightspace

We propose a new software-based method for drawing
two-dimensional layered scenes of the kind used for interactive picture editing
and as the basis of standards such as Adobe's Portable Document Format (PDF).
This method combines various known techniques from the 3D and 2D graphics
literature with our own work to produce a feasible algorithm, which has been
implemented in prototype.

Primitive methods are based upon the \emph{painter's model}: the objects are
rasterized individually from back-most to front-most. Later images overwrite
part or all of earlier objects where overlap occurs. When antialiasing or
partial transparency are present, each new object is combined with the
previously-rendered composite using the rules of \cite{porterduff} instead of
simple overwriting.

The disadvantage of this scheme is that objects which are partly or completely
obscured by those nearer the front are still calculated fully. This results in
wasted composition and antialiasing operations, as well as unnecessary
rasterization of parts of the object which will not affect the final composite.
Our method reduces the number of objects rendered, the portion of each which
need be rendered, and the recalculation required when a scene changes.  Scenes
in illustration graphics contrast sharply with those in real time
three-dimensional visualization. Since the emphasis in illustration is on
interactive editing, there tend to be many fewer objects, but each may take much
longer to rasterize---a large polygon or brush stroke with a high-radius
gaussian blur applied to it may take a significant portion of a second to
calculate. Much illustration graphics is destined for print use, at resolutions
of 100,000 pixels or more in each direction. Even on screen, the work increases
rapidly as monitors are available with much higher resolution than before. The
extra analysis required to minimize the amount of work to draw a scene (and
redraw it when it changes) by calculating only the part of each object which is
visible may therefore become useful. Moreover, scenes in illustration graphics are
often amenable to this kind of optimization since they tend to contain many
overlapping objects.  In an interactive editor, frame-to-frame coherence is high
because it is likely that just a few of the objects will be modified each time
the scene changes.  Since the user can only modify something once he has already
selected it, these modifications may be partly predicted. 

With more computing power, the ability to edit the document truly interactively
has been implemented in some programs. The scene updates as the user moves or
modifies an object, rather than displaying an outline or wire frame and updating
only when the change is certain. Due to the lack of a suitable rendering and
caching framework to minimise the work required at each frame, this is often
slower than it ought to be and not implemented for all types of scene
modification. We can not assume there is a limit to the number or type of
objects we might want to be rendered in real time---give an illustrator a new
object, and he will use a dozen on top of one another for an effect unforeseen
by the programmer.

The claimed efficiency of our hidden-surface removal algorithm is predicated
upon the following hunch:

\begin{quote}

\emph{When very high-quality rendering is used or the scene contains very
complicated objects, the cost of calculating the set of pixels affected by an
object is insignificant when compared with the cost of rendering the object.}

\end{quote}

Our intuition is that this is likely to be true for more complicated objects such as brush strokes, or
objects with fancy fills or effects like blurring.

In Section~\ref{litreview} we look at other work in this field.
Section~\ref{mainsection} is our major contribution---a new algorithm for
rendering two-dimensional scenes. We describe the data structures and how
they apply to simple objects such as polygons and brush strokes.  We exhibit a
hidden surface removal algorithm which ensures that only those pixels of an
object's rasterization whose values contribute to the final image will be
calculated. In Section~\ref{changes} we describe how to find the minimal portion
of an image which needs to be redrawn when the scene changes.  We discuss the
addition of a caching mechanism to exploit coherence between successive frames
of an animation or interactive session.  In Section~\ref{antialiasing} we
describe antialiasing in our system, and show how to modify the renderer to
address efficiently the problem of \emph{correlated mattes}, a characteristic
defect in the painter's model. In Section~\ref{filters} we add a system of
primitive-combiners called \emph{filters} providing both vector and raster
effects. We show how the efficiency of hidden surface removal holds over
filters. Finally, we conclude and suggest further work.

\section{Related Work in Computer Graphics}
\label{litreview}

Our method, while it has some novelty, uses work from a wide range of previous
work, both in two- and three-dimensional graphics.

The simplest way to render a two-dimensional scene is to render each layer,
retaining transparency information, and then to compose the layers one at a time
using the methods described in \cite{porterduff}. A good introduction can be
found in \cite{Smith95alphaand} and \cite{smithcomposit}. Hidden surface removal
in two dimensions is a special case of three-dimensional hidden surface removal
where we already have a complete depth ordering on the objects to be drawn, but
do not yet know whether we need draw all of each object, since objects may
overlap partially or completely.

Traditionally, two-dimensional systems have not considered the problem of
correlated mattes (where multiple partially overlapping or intersecting objects
contributing to a pixel cause wrong results). We extend our system to handle
this in Section~\ref{correlated}. Two three-dimensional systems which deal with
this problem properly are Catmull's Pixel Integrator \cite{catmullhidden} and
Carpenter's A-buffer \cite{808585}. We reverse the rendering order (as is often
done in 3D graphics, and in 2D in \cite{Willis99}), and show how the number of
pixels requiring special attention is thus reduced.

Our system solves the hidden surface problem for two-dimensional layered scenes
allowing for an arbitrary antialiasing filter---not just a box filter as is
common---calculating pixel values for each layer only when needed (including
solving the problem of correlated mattes exactly, and only, when needed). This
improves on current systems which calculate the whole of each layer even when
that layer will be partially or completely covered by an object further forward
in the scene. We do not discuss the mathematical foundations of antialiasing
theory, but for a description of the box filter's insufficiency, see
\cite{smithpixel}.

The success of our system relies upon the storage of fragments of rendered
content (\emph{sprites}) and sets of pixel locations (\emph{shapes}) being
efficient in the presence of plain fills, fancy fills and complicated
antialiased pixels. These concepts are introduced in the context of the
composition of primarily bitmap graphics in \cite{smithsprite}. We recast those
into the domain of a vector graphics editor, taking advantage of the
preponderance of non-rectangular objects. It also requires that set-based
operations on shapes are fast, and that composition of sprites is fast. We use
data structures similar to Wallace's cartoon cel work \cite{806813} and
Froumentin \& Willis' IRCS \cite{Willis99}.

Our system of primitive combiners called \emph{filters} is rather like that
developed by \cite{Bier1993}. Our method of drawing brush strokes (which we use
as one example of a non-polygon primitive) comes from \cite{Whitted1983}.

A discussion of some of the issues involved in writing an interactive editor for
two-dimensional scenes such as ours is in \cite{Fekete1996}.

\section{Hidden Surface Removal}
\label{mainsection}

A \emph{scene} consists of a number of objects, ordered by depth.We allow the
raster representation of an object to depend upon objects further to the back of
the scene, but it must be independent of those nearer the front. This allows for
our new \emph{filter} objects (see Section \ref{filters}). Most objects
(including partially transparent ones) do not depend upon objects nearer to the
back.

In order to be able to calculate the minimum work required to render a scene, it
is necessary to retain the essence of an object's geometry when it is
rasterized. We define a set of data structures lying somewhere on the boundary
between the geometric and rasterized planes.

\begin{figure} \centerline{\includegraphics[width=8.95cm]{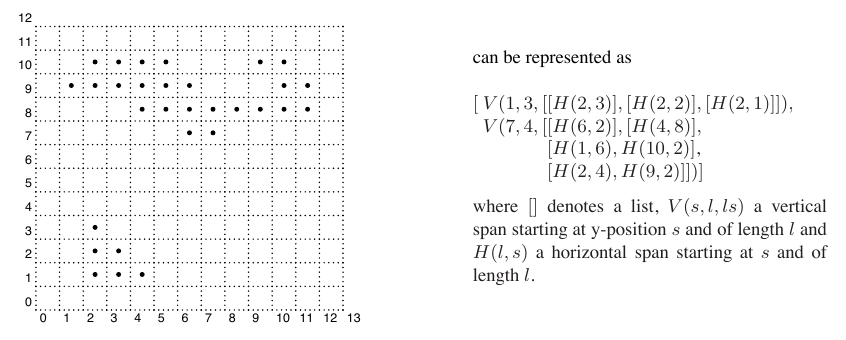}}

\caption{Possible raster set representation for shapes. For a sprite, the same
structure is used, but with pixel data added to each horizontal span, possibly
compressed with run length encoding.}    

\label{setrep}
\end{figure}

The \emph{shape} of an object is the set of pixel positions whose members are at
least those pixels which are expected to be not-wholly-transparent in its
rasterized representation.  The efficiency of the algorithm relies upon this
being close to the minimal set, but its correctness does not. The set is likely
to be minimal for objects with simple geometries (such as polygons) but not for
more complicated ones (such as point clouds, implicit geometries or objects
processed in highly non-linear ways). The set is stored in a spatial data
structure; the current scanline-based implementation is described in
Figure~\ref{setrep}. 

\begin{figure}
\centerline{\includegraphics[width=8.8cm]{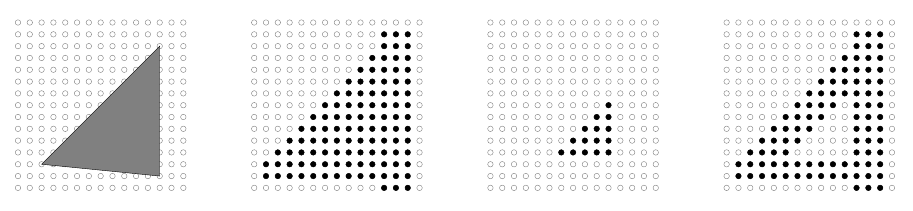}}

\caption{From left to right: A polygon in vector space and its shape, minshape
and maxshape as sets of pixels (filled) when using an antialiasing filter with
square footprint twice the interpixel spacing.}    

  \label{fig:shapes}
\end{figure}

The \emph{minshape} of an object is the set of all coordinates of pixels where
the object influences the pixel completely i.e. for which the geometry does not
alter the value of the function representing the fill colour of the object; for
a polygon this is all pixels for which the antialiasing footprint is contained
entirely within the polygon.  For very complicated geometries (for instance
particle clouds) the minshape is likely to be the empty set. The \emph{maxshape}
of an object is the set of pixels in its shape but not in its minshape. Clearly
only one of the maxshape and minshape need be stored, the other being derived
when required by set difference. A polygon and its shapes are shown in
Figure~\ref{fig:shapes}.

The \emph{sprite} of an object is its (partial) rasterized representation,
providing a set of colour values corresponding to some or all of the coordinates
in its shape, depending how much has required rendering to this point.

\begin{figure}
\centerline{\includegraphics[width=9cm]{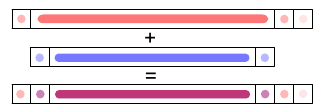}}

\caption{Composing a line of two partially transparent sprites, preserving
run-length encoded spans.}

\label{fig:compositing}
\end{figure}

\paragraph*{Calculating sets for Polygons}
\label{setsforpolygons}

Here we give some detail of the calculation of shapes and sprites for a
primitive such as the polygon---the process is a little different from
traditional polygon rasterization. The shape of a polygon contains every
coordinate which will have a not-wholly-transparent value under the rasterizing
scheme used to calculate its sprite. This is a simple extension to the standard
method for rasterizing filled polygons (\cite{Foley96}, p92). Scanlines are
considered to have a finite width equal to the diameter of the footprint of the
antialiasing filter rather than as a zero-width line. The edge list techniques
in \cite{Foley96} generalize simply.

\begin{figure}
\centerline{\includegraphics[width=8.7cm]{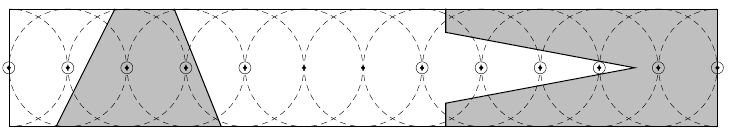}}

\caption{A finite-width scanline and part of a polygon crossing it (shaded),
together with the calculated shape scanline (marked as small circles around
pixel centres for those pixels included in the shape). The large, dotted
overlapping circles are antialiasing filter footprints, though we approximate
them with squares for this process. If the left-most pixel here is pixel $0$,
$spans(t)$ is {\small\{(1--3, 7--12\}}, $spans(b)$ is {\small\{0--3, 7--12\}}
and $covered(e)$ is {\small\{0--4,7--12\}} so the result is {\small\{0--4,
7--12\}}.}    

\label{fig:polygonshape}
\end{figure}

Consider a scanline and the list of edges which intersect it, as illustrated in
Figure~\ref{fig:polygonshape}. The situation is more complicated than for a zero
width scanline. The spans generated by the top and bottom scanline edges may be
different, and there may be edges lying partially or wholly within the scanline
that must contribute. Let $\mathit{spans}(x)$ be the sequence of spans derived
from the list of edge crossings $x$. Let $\mathit{covered}(x)$ be the sequence
of spans composed of the coordinates whose associated footprint is intersected
by one or more of the current edges $x$ which cross neither the top nor bottom
edges.  Then, if the current edge list is $e$, the list of crossings at the top
edge $t$ and those at the bottom $b$, the spans are
$\mathit{spans}(t)\cup\mathit{spans}(b)\cup\mathit{covered}(e)$ where $\cup$
combines touching or overlapping spans to form the minimal set.

When an object is to be rasterized to a sprite, the required shape will have
been calculated by the renderer by intersecting the object's shape with the
update shape (the set of pixels in the scene which have been determined to
require rendering). The job of the rasterizer, then, is to generate a partial
sprite of the same shape. Plain fill types, where the fill colour is constant
rather than dependent upon the coordinates of the pixel are treated
differently---a run length encoded subspan is generated for runs of pixels which
all have no edges intersecting their filter footprint. The antialiased parts
around the edges of the polygon are generated as usual.

Figure~\ref{fig:compositing} shows how run length encoding of polygon scanlines
is preserved over composition in our spatial representation of sprites. This
encoding allows spatial coherence to be preserved when plain shaded
possibly-translucent objects are composited over one another.

\paragraph*{Hidden Surface Removal Algorithm}
\label{hiddensurfaceremoval}

A classical painter's method renderer operates as follows. Objects are dealt
with from back-most to front-most. If an object intersects the update rectangle,
it is clipped, rendered and composited into a rectangular buffer. In this
method, objects contributing nothing to the final image are still drawn, and
objects which are partly obscured by ones in front are drawn in their entirety
(save for clipping to the update rectangle). This results in many wasted
calculations in polygon rasterization, antialiasing and compositing. It also
rules out the sensible use of caching techniques for interactive and animation
work, since the rasterized objects are very large. One possible approach is to
geometrically clip all the objects against one another, but this breaks down in
the presence of antialiasing and arbitrary primitives---how can one clip a brush
stroke against a point-cloud against a polygon?

To calculate only the parts of an object's rasterization which will contribute
to the final image, the order of rendering is reversed, considering the objects
front-most to back-most. Front-to-back rendering is obviously suitable for
entirely opaque objects, since the final pixel depends only upon the front-most
object affecting it. When objects are partially transparent (or,
equivalently\footnote{Most methods for antialiasing convert geometry information
  into a single coverage value for each pixel, so a half-covered opaque red
pixel is indistinguishable from a fully-covered half-transparent red pixel. See
Section~\ref{correlated} for a fuller discussion.}, antialiased), the operands
to the compositing function may usually be swapped, and the composite calculated
starting with the front-most object in the scene. When an object's rasterization
depends upon the objects behind it (such as an object with fill type `magnify
whatever is below by two, convolving it with a gaussian blur of radius five'),
extra work is required---we discuss this in Section~\ref{filters}.

The following description of the rendering process is independent of the
implementation of spatial data structures for sprites and shapes, the particular
primitives in use and their rasterization methods.

It is convenient to describe the rendering process in terms of a number of
fundamental operations on shapes and sprites. An efficient implementation may
not create all the intermediate structures suggested by this description.

\begin{itemize}
  \item $a \wedge b$, the set intersection of two shapes.

  \item $a \setminus b$, those pixels in shape $a$ which are not in shape $b$.

  \item $a \vee b$, the set union of two shapes.

  \item $a \circ b$, which composes sprite $b$ under sprite $a$, returning the
  composite together with a shape containing those members of $b$ whose
  analogous pixels in the composite are opaque (and so `finished').

\end{itemize}

\begin{figure*}
\centerline{\includegraphics[width=18cm]{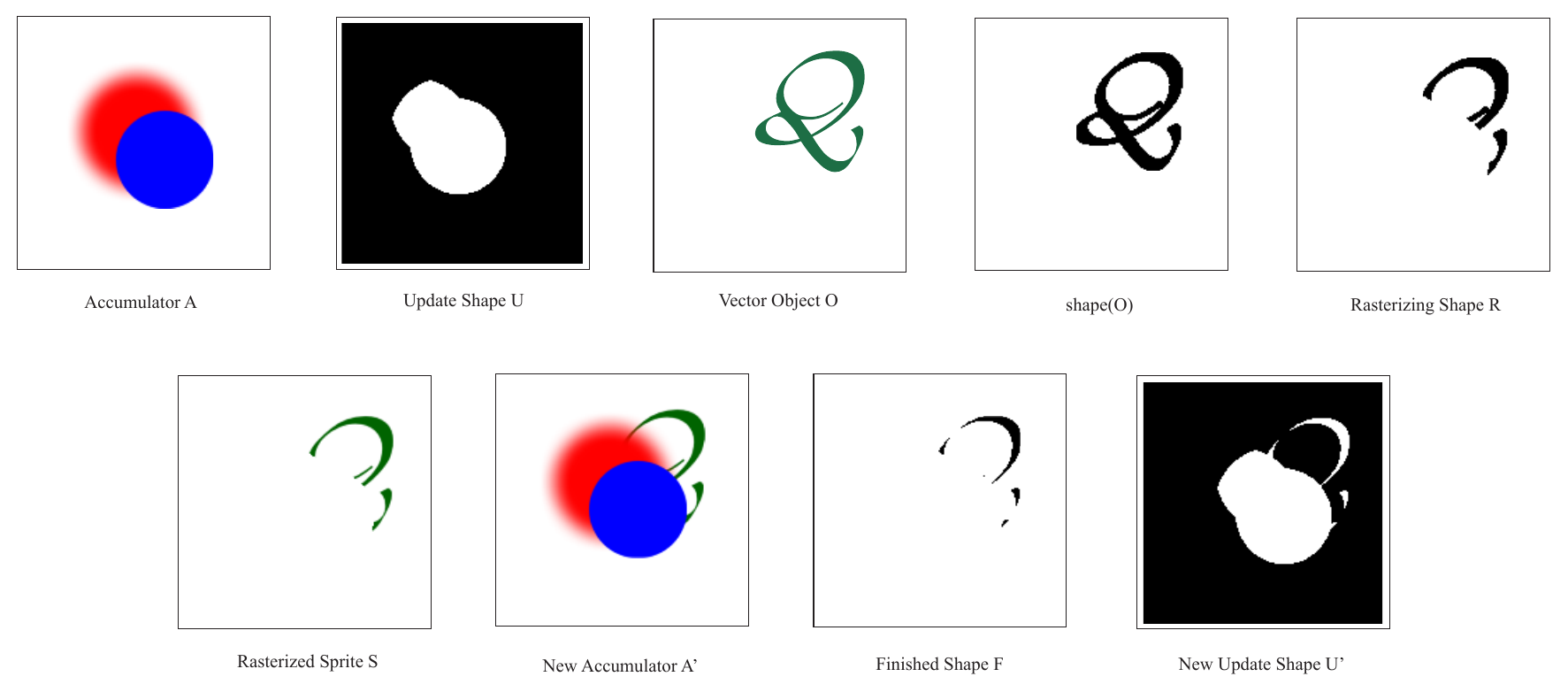}}

\caption{One stage of hidden surface removal front to back. The stages
illustrated are repeated for each object in the stack or until the update shape
is empty, starting with an update shape which covers the whole update region (a
simple rectangle, here).}

\label{fig:hiddensurfaceremoval}
\end{figure*}

One stage of this process is illustrated in
Figure~\ref{fig:hiddensurfaceremoval}.  Call the shape remaining to be updated
at this point (the set of pixels not yet in their final form) U, the current
object O, and the rendered result A (the \emph{accumulator}). The function
$\mathit{shape}(O)$ calculates the shape of an object $O$ and
$\mathit{rasterize}(O, R)$ calculates a partial sprite of $O$ for shape $R$.
Proceed as follows:

\begin{enumerate}

  \item Calculate the intersection between the update shape and the shape of
  the current object. Call this the \textit{rasterizing shape}, $\mathrm{R}$.
  \[\mathrm{R}=\mathrm{U} \wedge \textit{shape}(\mathrm{O})\]
  \item Rasterize the partial sprite corresponding to $\mathrm{R}$. Call this
  the \textit{object sprite} $\mathrm{S}$.
  \[\mathrm{S}=\textit{rasterize}(\mathrm{O},\mathrm{R})\]
  \item Compose $\mathrm{S}$ under the accumulator and find the pixels newly
  \emph{finished}.  A finished pixel is one which is entirely opaque and so
  cannot be affected by any more objects. Call this shape $\mathrm{F}$. The new
  accumulator is $\mathrm{A'}$.
  \[(\mathrm{A}',\mathrm{F}) = \mathrm{A}\circ\mathrm{S}\]
  \item The new update shape $\mathrm{U}'$ is $\mathrm{U}$ with all the newly
  finished pixels removed.
  \[\mathrm{U}' = \mathrm{U}\setminus\mathrm{F}\]
  \item Next object with $\mathrm{A} = \mathrm{A}', \mathrm{U} = \mathrm{U}'$.
\end{enumerate}

After all objects have been processed, the final accumulator sprite
$\mathrm{A}'$ has a shape which is a subset of the original $\mathrm{U}$. If the
output is to be used for rendering to a device which does not have an alpha
channel (for instance, a screen), the last object in the scene will be the
\emph{background} (which is everywhere-opaque), so the final result will have
the same shape as the original U.

\section{Frame-to-frame Coherence}

\label{changes}
 
\paragraph*{Finding the Update Shape}

The rendering model for multiple frames is stateless; it is the job of the
program calling the renderer to decide the region which needs to be updated when
a change occurs.

\begin{figure}
  \centerline{\includegraphics[width=8.7cm]{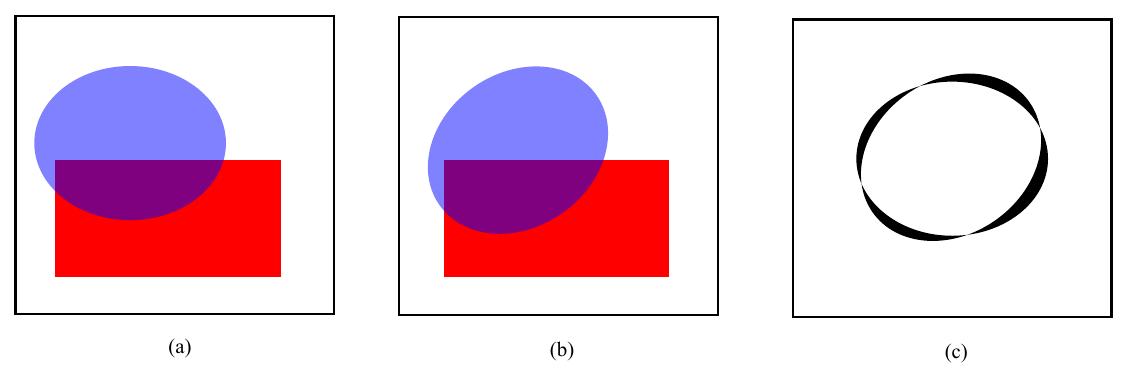}}
  \caption{Finding the update shape. The blue oval has been rotated between
  frames (a) and (b). The update shape is shown in (c). The pixels in the centre
of the oval have not changed colour so do not need to be redrawn.}
\label{fig:coherence} \end{figure}

Call the set of pixels whose values are considered to have changed the
\emph{update shape}. It depends upon the kind of operation (translation,
rotation, deletion etc.) that is performed, and upon properties of the object's
rasterization. When an object undergoes a rotation, for example, the update
shape is $a\vee a'$ (where $a$ is the old shape, $a'$ the new). If, however, the
rasterization of that object is independent of the operation (for instance, a
plain fill is independent of rotation), the update shape is
\[(\textit{shape}(a)\setminus\textit{minshape}(a'))\vee
(\textit{shape}(a')\setminus\textit{minshape}(a)). \] The situation is
illustrated in Figure~\ref{fig:coherence}. Since the shapes are likely to be in
a cache (See Section~\ref{caching}), and in any case may be calculated quickly,
we expect this to be efficient.  There are some circumstances when the update
shape is empty even though the scene has changed. For instance if the changes do
not affect the current area of the scene viewable on screen, or if the operation
is known not to affect the rasterized representation of the scene (for example,
ungrouping a group of objects in an interactive illustration package).

All calculations determining the update shape requiring rasterization in a
particular update cycle are implicitly taken to be in intersection with the
\emph{region of interest}, for example the viewport of the current window.

\paragraph*{Caching for Interactive Changes}

\label{caching}

An important side effect of exact hidden surface removal is that the size in
memory of an object's sprite is likely to be smaller due to the use of a spatial
data structure and the fact that many objects will only be partially rasterized.
This makes caching of part or all of the rasterized data feasible so that, when
the scene changes, parts of sprites which have not changed need not be
recalculated (when needed as part of a changed composition). The cache can store
partial or complete sprites corresponding to the rendered portions of each
object. We also store shapes (the sets representing the pixels affected by an
object), since these are small and frequently required for the calculations in
the hidden surface algorithm.

When a new item is added to the cache, or an object's partial sprite in the
cache extended, one or more cache items may need to be removed to make space.
To decide which to remove, the cache items are scored according to various
metrics (how recently the item was last used, the time the object took to
render, the size in memory of the item, the type of the item -- shape or
sprite). Often, it is useful to keep previous generations of an object in the
cache too---a frequent operation in interactive graphics programs is undo, which
should be fast. This also means that, upon zooming in on a part of the scene to
inspect it and then returning the original scale, the scene should not have to
be recalculated.

It may prove desirable to cache precomposited portions of the image to prevent
having to recompose at every update. This works due to \emph{depth coherence} in
successive editing operations. That is, when the user selects an object, it is
likely they are about to modify it.  One composite sprite for all the objects
below the back-most currently-selected object can be made so that as it is
(interactively) modified, the re-rendering of the scene is faster.

When the user is interactively, say, rotating an object, the screen must be
updated in as near real-time as possible, but the change is not committed to the
scene until the mouse button is released. Typically the user can also cancel the
interactive operation by pressing the escape key. This scheme has implications
for efficient use of the cache. Caching each of the dozens of generations of the
sprite of an object which result as the interactive modification is made is
unacceptable. Consider how to react when the user has his chosen result and
wishes to commit the modification, or wishes to abandon the modification. Upon
abandonment, the renderer is called with the update shape $a\vee a'$ (where $a$
is the old shape, $a'$ the latest of the interactive shapes) and the old scene.
Upon commit, the situation is rather more complicated. The new shape and sprite
have just been calculated so it is important to avoid recalculating them. This
is done by keeping the latest rendered object(s) in the interactive change in a
private single-entry cache, moving them into the main cache upon commit.

One of the most common operations when editing a scene is translating an object.
Since this operation is often done with the mouse, in a large proportion of
cases the translations in $x$ and $y$ will be integers at the current viewing
transform. This means the rasterized representation does not change. The
geometry of the object changes, but its shape and sprite can just be translated.
This does not apply to all objects---filters (see Section \ref{filters}), for
example, can change their rasterized representation based upon their position.

\section{Antialiasing and the Problem of\\ Correlated Mattes}
\label{antialiasing}

Antialiasing in our system is achieved by integrating over a filter. In our
particular implementation, we use a gaussian with a filter footprint width equal
to twice the interpixel spacing. Quickly calculating this integration is
discussed in \cite{Feibush:1980:STU:965105.807507},
\cite{Catmull:1984:AVS:964965.808586} and \cite{duffpolygon}.

\begin{figure}
 \centerline{\includegraphics[width=9cm]{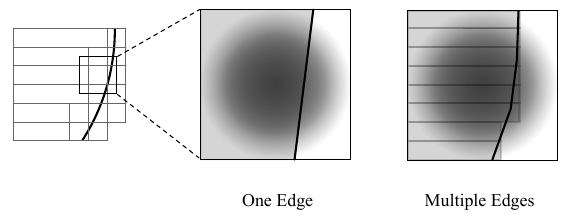}}

\caption{The calculation of antialiasing for polygons using an arbitrary filter.
  When there is just one edge crossing the filter, a lookup table for filter
  values based on the two endpoints and inside-outside nature of the top left
  point is sufficient. When there are multiple edges, we must do the integration
in sections, by calculating the shape spans and looking these (start x, start y,
length) spans up in another pre-calculated table.}

\label{fig:antialiasingfilter}
\end{figure}

The coverage calculation can be done from the same edge lists which were used to
calculate the polygon's shape. In the simplest case there is just one edge
crossing the filter footprint and the volume under the filter may be looked up
in a simple table mapping the triple (start point, end point, whether top left
corner of footprint is inside or outside) to a value. For reasonable subpixel
granularity, this lookup table is small---especially if the fourfold symmetry
of the filter is exploited to reduce its size. In our system, the table is just
a few hundred bytes for a 16x16 subpixel arrangement. When there are multiple
edges, the set of subpixels representing the rasterized edge within the filter
footprint is calculated using almost the same system as for calculating shapes
given earlier, scaled up so there are multiple spans per pixel. Each subspan can
then be looked up in a table mapping (start x, start y, length) to integrated
values, which are then summed to find the total contribution.  Figure
\ref{fig:antialiasingfilter} shows an example.

\label{correlated}

When a shape is rasterized using antialiasing, geometry is exchanged for a
single coverage value, and information is lost.  This manifests itself when such
images are composited with one another, since it is not known how much of the
top object obscures the bottom object in each pixel. Porter and Duff [1984] cite
one of the worst cases, where the same object is used twice in a compositing
expression---it covers itself exactly, but the algorithm blends the colours
nonetheless. 

\begin{figure}
 \centerline{\includegraphics[width=6cm]{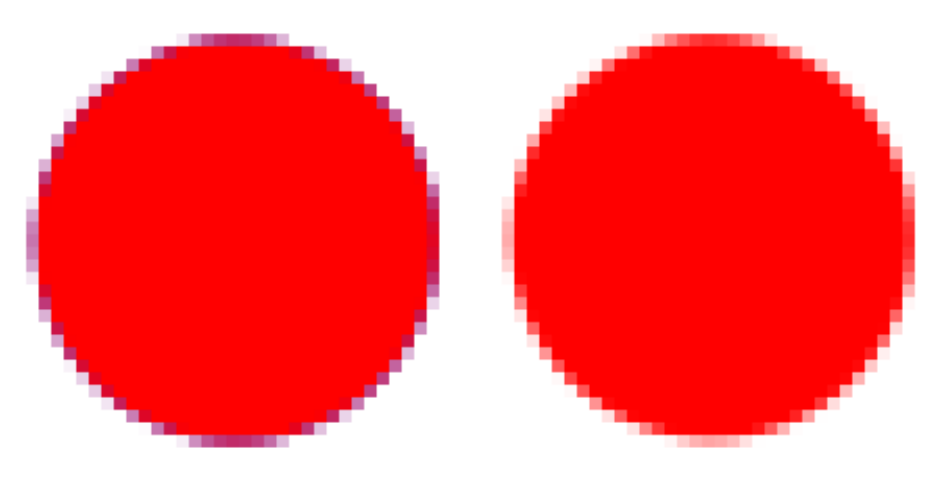}}

\caption{The problem of correlated mattes. A red circle placed exactly over a
blue one. The left pair composited as usual; the right pair with proper
treatment of correlated mattes.}

\label{fig:correl}
\end{figure}

Consider Figure~\ref{fig:correl}. Since the red entirely covers the blue, no
blue should show through. However, rasterization exchanged geometry for
coverage, so the renderer has no information to decide if the blue object is a
partially transparent one covering the whole of the pixel, or an opaque one
covering part of it. It has no choice but to assume the colours blend. This is
known as \emph{the problem of correlated mattes}. Two other manifestations of
this problem are visually disturbing: When a thin line appears to darken as it
crosses other lines or itself, and when two abutting polygons appear to have a
thin line separating them. Commercial illustration graphics programs have
traditionally ignored this problem.

Any method to solve the problem of correlated mattes will involve a hidden
surface algorithm of some kind at certain pixels. Two classic solutions are
Carpenter's \emph{A-buffer} \cite{808585} and Catmull's \emph{Pixel Integrator}
\cite{catmullhidden}. The A-buffer keeps a bitmapped buffer for each pixel
affected by polygon edges. The Pixel Integrator analytically clips each polygon
to a square surrounding the pixel, and computes each polygon's visibility
exactly.

\begin{figure}
\centerline{\includegraphics[width=4cm]{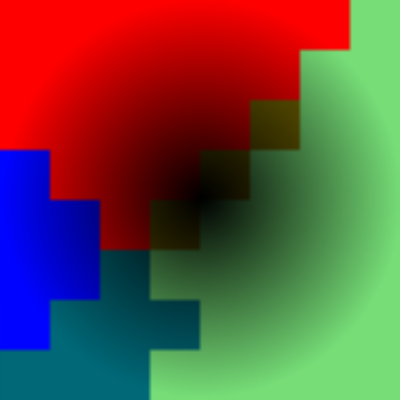}}

\caption{Solution to the problem of correlated mattes by using our usual
front-to-back hidden surface removal process within the antialiasing filter
footprint of a single pixel. The results are then weighted by the filter to
produce a single result, once the pixel is finished.}

\label{fig:correlatedmattes2}
\end{figure}

We should only perform hidden surface removal within individual pixels when it
would make a difference to the final composite. It is quite simple to extend our
current system with an optional subpixel analysis at the cost of slower
rendering: pixels forming part of the maxshape of an object are initially
represented by a square matrix of subpixels, covering the footprint of the
antialiasing filter in use and stored as a sprite (so space-efficient).  When
compositing into the accumulator, subpixels are composited with one another in
the usual manner. A pixel is finished when all its subpixels are finished. At
this stage, the antialiasing filter is applied and the representation of that
pixel reverts to normal. This ensures that no more work is done than is
required. When all objects have been rendered, any remaining
subpixel-represented pixels in the accumulator are normalized.
Figure~\ref{fig:correlatedmattes2} shows an example single pixel accumulator
under this system. Note that the shapes are rendered in this subpixel system
without antialiasing, so that in the case two identical polygons on top of one
another, none of the one beneath would show, as required.

%This system is slower than normal rendering, but may be turned on when
%highly accurate rendering is required. It can be restricted to some objects
%only---for example, we may apply it to polygons, but consider representing the
%colour-falloff of a wide, blurred brush stroke at the subpixel level to be
%excessive. 

\section{Filters}
\label{filters}

Sometimes it is useful for the object's sprite to depend upon the rasterized
composite of the objects below it, or for the object to be able to remove
objects or change their geometry whilst they lie under it. We call this kind of
object a \emph{filter}. The concept is not new (see, for example,
\cite{Bier1993}), but our system is more general than previous
efforts: filters can read from directly underneath themselves, or from other
parts of the scene, modifying the result either geometrically or using raster
effects. Filters can be made from any type of primitive. For instance, the
filter `blur the scene below' requires rendering the scene below normally, and
then blurring it to form the sprite of the filter object.  The filter type `make
all objects below have thin lines and transparent fills' (an outline or wire
frame effect) involves modifying the object geometries themselves. A filter has
an opacity (plain, or varying) just like any other object, representing the
extent to which the filter affects each pixel.

A filter represents an interruption in the usual front-to-back rendering order.
Filters complicate the rendering process significantly, especially if we are to
preserve the efficiency gains we have seen with simple front-to-back rendering.

A filter consists of:

\begin{itemize}

\item The \emph{geometry} of the filter, which is a primitive of some kind
  (polygon, brush stroke etc).  Only the alpha channel is used, for defining
  where and to what extent the filter acts. The rasterizations of the filter and
  base scene will be blended in proportion to this geometry, allowing for
  correct antialiasing and partially transparent filters. 

\item The \emph{scene function}, which takes as input the current scene under
  the filter and the shape representing the part of the filter's shape which
  requires rasterization. It returns a modified scene (perhaps removing, adding
  or changing objects) whose rasterization will be required before the filter
  itself can be rasterized, a shape in which that rasterization is required (the
  \emph{reading shape}), and a shape representing how much of the filter itself
  need be rendered (the same or a subset of the input shape).

\item The \emph{filter function}, which performs computation on the sprite
  returned by the rendering of that part of the scene which is returned by the
  scene function. For instance, it may blur it.

\item The \emph{update function}, which is used to determine if areas of the
  filter need updating when some other area of the scene underneath needs
  updating.

\end{itemize}

The scene function can, of course, produce a scene with more filters in it,
leaving open the possibility of a non-terminating renderer. An interactive
graphics program using the renderer must prevent this by construction.

\paragraph{Rendering with Filters}

\begin{figure}
\centerline{\includegraphics[width=8.5cm]{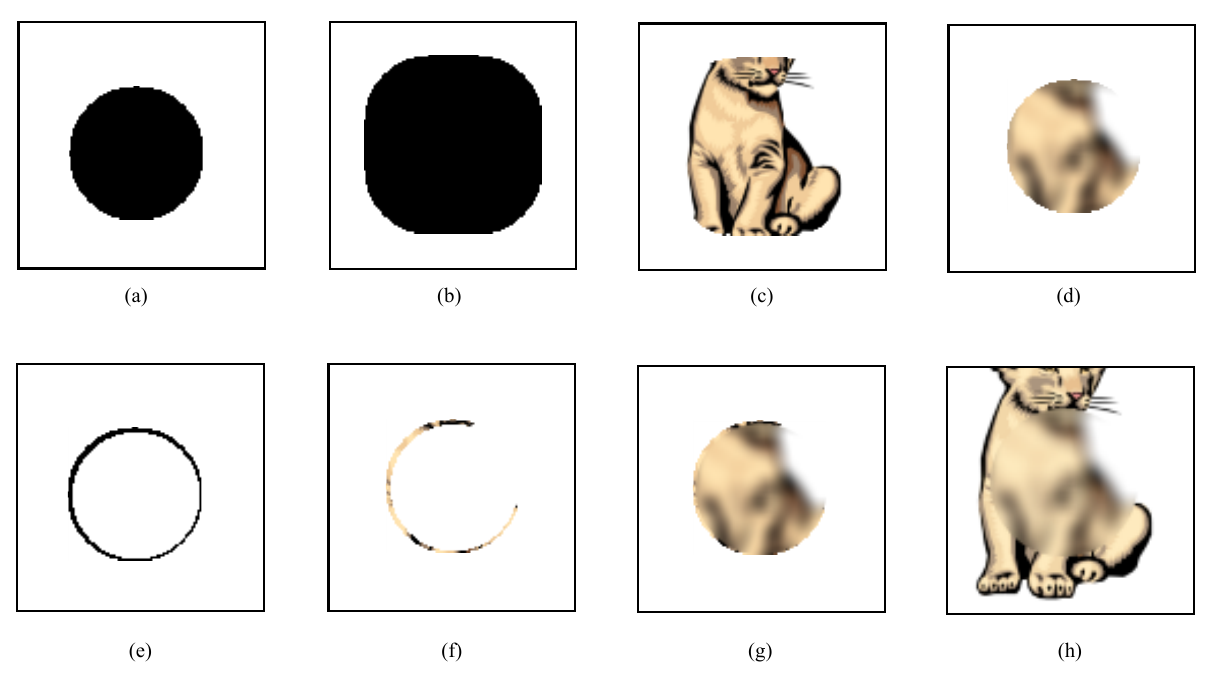}}

\caption{Rendering a filter using its \textit{geometry}, \textit{scene function}
  and \textit{filter function}. Here the geometry is a polygon, the scene
  function expands the reading shape (to provide enough data for the blur to be
  rendered) but does not alter the scene, and the filter function blurs the
  rendered scene. (a)~the shape of the filter's geometry (b)~the reading shape
  (c)~the scene rendered in the reading shape (d)~the result of the filter
  function (e)~the pixels where the filtered and unfiltered scenes must be
combined (here just the antialiased pixels) (f)~the original scene rendered in
those pixels (g)~sprites (f)~and~(d) combined using the Porter Duff 'plus'
operator. (h)~the whole scene including the filter.}    

\label{fig:filter}
\end{figure}

When a filter is encountered, various of the basic set operations defined above
are used, together with the filter's specification, to calculate its sprite and
finished pixels. We proceed as follows (letters from Figure \ref{fig:filter}):

\begin{enumerate}
  \item Picture (a) is the geometry of the filter. Use the filter's scene
  function to find the reading shape (b) and the modified scene.

  \item Render (c) the modified scene in the region of the filter which we are
  rasterizing. Execute the filter function to make (d).

  \item Find the pixels (e) in the rendered alpha channel of the filter
  geometry which are not opaque (this is the maxshape of the filter geometry).

  \item Rasterize (f) the original scene in those pixels.

  \item Blend (g) the original and filter together, attenuating the filter in
    proportion to (and the original scene in proportion to the complement of)
    the alpha channel of the filter geometry and combining them where they
    overlap using the Porter-Duff `plus' operator. Picture (h) shows this in the context of the eventual, fully-rendered image.

  \item The finished pixels for the filter object are all those in the shape of
    the filter's sprite, rather than those which are actually opaque as for a
    non-filter object. This allows filters which take paint away from the canvas
    to function correctly.  \end{enumerate}

\paragraph*{Some Example Filters}

\begin{figure}
\centerline{\includegraphics[width=9.2cm]{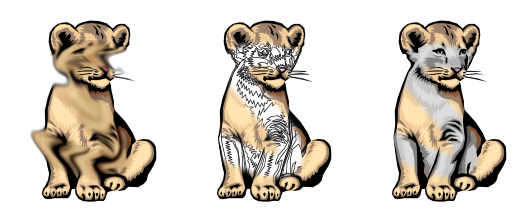}}

\caption{Filter examples, using a brush stroke as the filter geometry. From left
to right: smear along the path of the brush stroke, outline (wire frame),
monochrome.}    

\label{fig:filterexamples}
\end{figure}

\begin{figure*}
\centerline{\includegraphics[width=12.1cm]{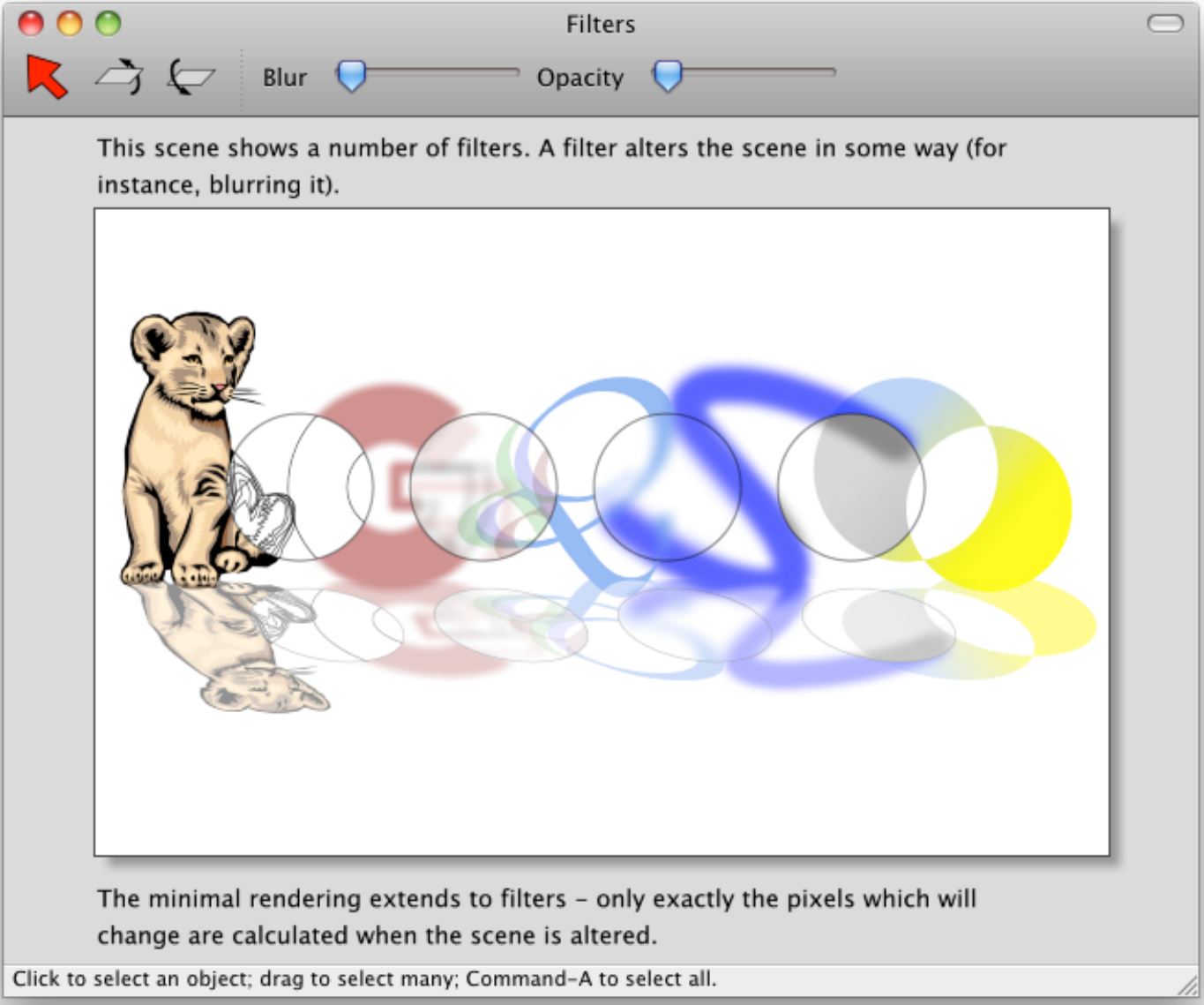}}

\caption{A complete scene shown in our basic editor. The scene shows five
  objects with four filters in front of them. A final filter on top of the scene
  reflects, shears and fades the whole scene to form a reflection. The filters
  (from left to right) are outline (wire frame), colour splitting, blurring and
monochrome.}    

\label{fig:examples}
\end{figure*}

Here we describe some filters to illustrate how different scene and filter
functions are used. Figures~\ref{fig:filterexamples} and~\ref{fig:examples} show
the effect of some other filters, with various geometries.

\begin{figure}
\centerline{\includegraphics[width=8.7cm]{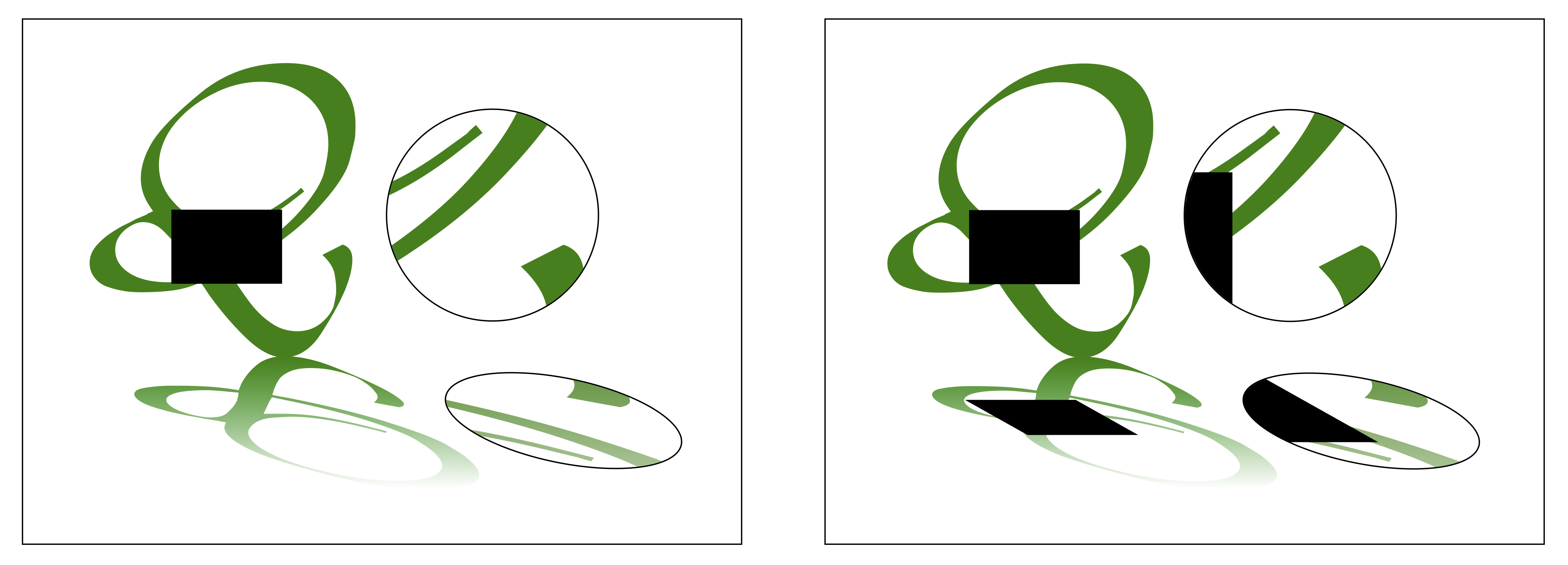}}

\caption{Frame-to-frame coherence in the presence of filters. The left-hand
  image shows a scene with a zoom filter overlaid with a shadow filter. The
  black rectangle is the region of the scene requiring update for whatever
  reason. In the right hand image, the filters' update functions have been used
to propagate the update shape through the scene to determine which other regions
have been invalidated by the change.}    

\label{fig:filtercoherence}
\end{figure}

\begin{itemize}

\item \textbf{Blur} To blur the scene in a given area. The scene function
  returns the scene under the filter unaltered, the reading shape is the shape
  of the filter geometry convolved with a rectangle of the width and height of
  the convolution kernel to include the extra pixels required for calculation of
  a blur. The filter function convolves the sprite by the kernel in the region
  of the filter geometry, returning a sprite of just those pixels in the
  original filter geometry.

\item \textbf{Cutting a hole} To cut a hole in the entire scene below the
  filter. The scene function returns the empty scene, the reading shape is the
  shape of the filter geometry and the filter function is the identity function.
  A hole can be cut just through one object by returning the scene with that
  object removed.

\item \textbf{Affine transform} To produce a magnify, reflect or similar effect.
  The scene function applies some affine transform to the scene. The reading
  shape is unaltered. The filter function is the identity function, or some
  function such as blurring or tinting, avoiding the needless composition of
  several filters when one could suffice. 

\end{itemize}

When an object is altered, added or removed from the scene, an initial update
shape is calculated using the methods already described. When the scene contains
filters, part or all of each filter lying outside this region may need to be
recalculated. A simple but inefficient method would be to invalidate the
entirety of all filters when a change is made to the scene.  However, we wish to
preserve the efficiency of exact hidden surface removal over filters, making
them almost as efficient as basic shapes. We associate an \emph{update function}
with each filter. The final update shape is given by the composition of the
update functions of the filters from the back-most modified object forwards,
again taken in intersection with the region of interest. The process is
illustrated in Figure \ref{fig:filtercoherence}.

\section{Other Primitives}

\begin{figure}
\centerline{\includegraphics[width=9cm]{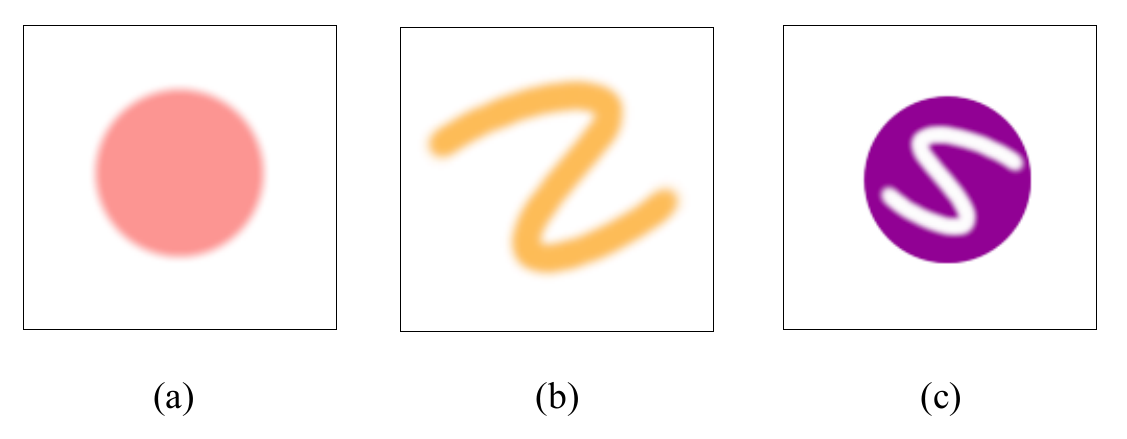}}

\caption{Examples of other primitives: (a)~a blurred polygon (b)~a brush stroke
and (c)~a polygon with a brush stroke shaped hole in it. Our system can perform
set operations on geometries to build new ones, retaining the efficiencies of
their shape and sprite representations.}

\label{fig:blurredobjectetc}
\end{figure}

Nothing in our hidden surface algorithm requires that polygons be used---any
primitive which can be rendered (i.e. its shape and sprite calculated for a
given region) may be used. There is no need to convert them to polygons---they
retain their own geometry. Figure~\ref{fig:blurredobjectetc} shows:

\begin{enumerate}[(a)]

\item A blurred polygon (our renderer allows any object to be blurred).

\item A brush stroke created with the techniques in \cite{Whitted1983}.

\item The subtraction of a brush stroke from a polygon, forming a brush stroke
  shaped hole in the polygon. Our renderer allows such set operations on
  any combination of types of geometry, using the same operations to calculate
  the shape and partial sprite efficiently.

\end{enumerate}

The geometry of a filter may likewise be formed of any primitive or combination
of primitives. For example, Figure~\ref{fig:filterexamples} uses brush stroke
shaped filters.

\section{Conclusion}

We have presented a method of hidden surface removal for static and moving
scenes containing arbitrary primitives and primitive-combiners. We have shown
how this works together with updated versions of classic antialiasing and
rendering methods to calculate information only when it is required.

Many systems nowadays have either specific accelerated graphics hardware, or
multiple CPU cores, or both. Whilst it is always useful to keep a reference
software implementation, could this system be implemented wholly inside graphics
hardware? To what extent might it be made parallel? More work is required to
define the best caching mechanisms based upon empirical observation. There is
plenty to be done on efficiency for particular kinds of scenes. It remains to
build the renderer into an interactive graphics application for insights into
its use in real situations, and to add low level optimisations to complement the
algorithmic efficiency already present.

\paragraph{Acknowledgments}The author is indebted to Alvy Ray Smith for his careful and constructive comments on an earlier draft of this work.

\bibliographystyle{acmsiggraph}
\nocite{*}
{\small\bibliography{coherence}}

\end{document}